\documentclass[12pt]{iopart}
\usepackage{graphicx}
\begin{document}

\title[Inertial and Gravitational Mass]
{Concepts of Intertial and Gravitational Mass}

\author{A. Widom}

\address{Physics Department, Northeastern University\\ 
110 Forsyth Street, Boston MA 02115\\
{\it Email:} allan.widom@gmail.com}

\begin{abstract}
The general relativistic notion of gravitational and inertial mass 
is discussed from the general viewpoint of the tidal forces implicit in 
the curvature and the Einstein field equations within ponderable matter. 
A simple yet rigorously general derivation is given for the Tolman 
gravitational mass viewpoint wherein the computation of gravitational 
mass requires both a rest energy contribution (the inertial mass) and 
a pressure contribution. The pressure contribution is extremely small 
under normal conditions which implies the equality of gravitational 
and inertial mass to a high degree of accuracy. However, the pressure 
contribution is substantial for conformal symmetric systems such as 
Maxwell radiation, whose constituent photons are massless. Implications 
of the Tolman mass for standard cosmology and standard high energy particle 
physics models are briefly explored.  
\end{abstract}

%\maketitle

Any notions of a gravitational mass as a source for gravitational fields and 
inertial mass in general relativity\cite{Einstein:1953} must derive their 
magnitude from the energy-pressure tensor \begin{math} T_{\mu \nu }  \end{math} 
of ponderable matter. The continuous material media vector velocity 
\begin{math} v^\mu  \end{math} and the scalar energy density  
\begin{math} \varepsilon \end{math} may be computed via the unique 
time-like eigenvector of the energy-pressure tensor. In detail, 
\begin{equation}
T_{\mu \nu }v^\nu =-\varepsilon v_\mu \ \ \ {\rm wherein}
\ \ \ v^\mu v_\mu =-c^2.
\label{gm1}
\end{equation}     
The energy-pressure tensor must then have the form 
\begin{equation}
T_{\mu \nu }=\varepsilon \left[\frac{v_\mu v_\nu}{c^2}\right]+P_{\mu \nu}
\ \ \ {\rm such\ that}\ \ \ P_{\mu \nu}v^\nu =0.
\label{gm2}
\end{equation}     
The anisotropic pressure tensor is thereby \begin{math} P_{\mu \nu} \end{math}, 
which on average yields the scalar pressure \begin{math} P \end{math} via 
\begin{equation}
P=\frac{1}{3}P^\mu _{\ \ \mu} \ \ \ {\rm equivalent\ to}
\ \ \ T\equiv T^\mu _{\ \ \mu }=3P-\varepsilon .
\label{gm3}
\end{equation}
In summary, 
\begin{equation}
T_{\mu \nu } =(\varepsilon +P)\left[\frac{v_\mu v_\nu}{c^2}\right]
+Pg_{\mu \nu}+\Pi_{\mu \nu},\ \ \ \Pi_{\mu \nu}v^\nu =0
\ \ \ {\rm and}\ \ \ \Pi^\mu_{\ \mu}=0,
\label{gm2a}
\end{equation}     
represents the most general form for the energy-pressure tensor. From the above 
information contained in the energy-pressure tensor, we may discuss the 
differences between inertial and gravitational mass.

The inertial mass density of ponderable matter may be equated to the energy density 
in the local Lorentz rest frame of the continuous media, i.e. 
\begin{equation}
\tilde{\rho}=\frac{\varepsilon}{c^2}\ \ \ \ \ \ \ \ ({\rm inertial\ mass\ density}).
\label{gm4}
\end{equation}
The definition of inertial mass density in Eq.(\ref{gm4}) is more or less obvious. 
For example, if a point particle were moving along a space-time path, then 
consistent with Eq.(\ref{gm4}) the inertial mass \begin{math} m \end{math} 
would be computed from the particle vector momentum \begin{math} p^\mu \end{math}
via the usual rule \begin{math}-p^\mu p_\mu =m^2c^2  \end{math}. For any finite 
inertial mass \begin{math} m  \end{math},  
the scalar mass density \begin{math} \tilde{\rho }  \end{math} definition in 
Eq.(\ref{gm4}) amounts to the Einstein {\em rest energy} formula 
\begin{math} m={\cal E}/c^2 \end{math} but in a general relativistic form.
 
The proper definition of the {\em gravitational} mass density 
\begin{math} \rho  \end{math} of ponderable matter presents a more subtle 
problem. Tolman\cite{Tolman:1934} studied the problem of how to define the 
observable gravitational mass \begin{math} M \end{math} of a spherically symmetric 
large massive body. The required mass appears as the 
gravitational radius  \begin{math} R_s=(2GM/c^2) \end{math} 
in the Schwartzchild metric solution to the Einstein field equations.   
It was concluded that the definition of the gravitational mass density within the 
spherical body required both the energy density \begin{math} \varepsilon  \end{math} 
and the pressure \begin{math} P \end{math} according to the rule 
\begin{equation}
\rho =\frac{\varepsilon+3P}{c^2}=\tilde{\rho}+\frac{3P}{c^2}
\ \ \ \ \ \ \ ({\rm gravitational\ mass\ density}).
\label{gm5}
\end{equation}
The difference between inertial mass density \begin{math} \tilde{\rho}  \end{math}
and gravitational mass density \begin{math} \rho  \end{math} is thereby due 
to the pressure \begin{math} P \end{math} as described in the above Eq.(\ref{gm5}). 

Let us pause to give a {\em simple} yet quite {\em general} proof of the Tolman 
gravitational mass density Eq.(\ref{gm5}) which is free of the specific spherical 
symmetries of the Schwartzchild metric. Within ponderable matter with a flow velocity 
\begin{math} v^\mu \end{math}, there will exist tidal forces due to the gravitational 
curvature. The tidal force tensor is well known\cite{Weber:1961} to be given by  
\begin{equation}
\Phi_{\lambda \sigma }=R_{\lambda \mu \sigma \nu }v^\mu v^\nu .
\label{gm6}
\end{equation} 
The trace of the tidal force tensor may be evaluated employing the Einstein field 
equations, 
\begin{equation}
R_{\mu \nu}=\frac{8\pi G}{c^4}\left\{T_{\mu \nu}-\frac{1}{2}g_{\mu \nu} T\right\},
\label{gm7}
\end{equation}
in the form 
\begin{equation}
\Phi^\lambda _{\ \lambda}=R_{\mu \nu }v^\mu v^\nu 
=\frac{4\pi G}{c^2}\left\{2T_{\mu \nu}\frac{v^\mu v^\nu }{c^2}+T\right\}.
\label{gm8}
\end{equation} 
Eqs.(\ref{gm1}), (\ref{gm3}) and (\ref{gm8}) imply the tidal force tensor trace 
\begin{equation}
\Phi^\lambda _{\ \lambda} =\frac{4\pi G}{c^2}(\varepsilon +3P)=4\pi G\rho .
\label{gm9}
\end{equation}
In that Eq.(\ref{gm9}) is merely the general relativistic version of 
the Newtonian gravitational field equation 
\begin{math} \nabla^2\Phi =4\pi G\rho  \end{math} 
in a very thinly disguised form, the proof of the Tolman gravitational mass 
density Eq,(\ref{gm5}) has been completed.

Since the relativistic stability of ponderable matter requires that 
\begin{math} 3P\le \varepsilon  \end{math}, we have an inequality between 
inertial and gravitational mass densities 
\begin{equation}
\rho \le 2\tilde{\rho }.
\label{gm10}
\end{equation}
We note in passing that a {\em cosmological} term, when added into the Einstein 
field equations, may be viewed as a uniform but {\em negative} pressure in 
Eq.(\ref{gm2a}). Negative pressure metastable states of matter are available in laboratories. 
Furthermore, \begin{math} \rho < 0  \end{math} is {\em not} forbidden by any 
known general relativistic theorem. 
For the ordinary stable continuous matter around us, the positive pressure contribution 
to the gravitational mass is extremely small so that 
\begin{math} \tilde{\rho }\approx \rho \end{math} holds true to a sufficient 
degree of accuracy. For the Maxwell radiation field, one has 
\begin{math} \varepsilon =3P  \end{math} 
so that \begin{math} \rho = 2\tilde{\rho }  \end{math} for a gas of photons. That 
the gravitational mass is twice the inertial mass holds true for all model systems in which 
the constitutive particles are massless. 

The \begin{math} \rho = 2\tilde{\rho }\end{math} result holds true for 
field theoretical models exhibiting conformal symmetry. For example, the massless 
gluons which help model strong interactions contribute an energy density 
and pressure related by \begin{math} \varepsilon=3P  \end{math}. 
If one were to build an inertial 
mass with say a glue-ball made up of massless constituent gluons, then one would  
also build up a gravitational glue-ball mass at twice the inertial mass value.  
But as stated above, the masses in our neighborhood obey 
\begin{math} \tilde{\rho}\approx \rho \end{math}. 
It is thereby evident that glue-balls contribute very little to the observed 
neighborhood gravitational masses. Let us consider this type of result in somewhat 
more detail.

In the standard model of matter\cite{Weinberg:1996}, one begins with an 
\begin{math} SU(3)\times SU_{\rm left}(2)\times U(1)  \end{math}
field theory with conformal symmetry even for the quark and lepton 
sectors of the theory. The conformal symmetry is broken by a conjectured Higgs 
field which grows masses on some of the elementary particles, specifically   
\begin{math} (Z,W^\pm ,e,\mu ,\tau ) \end{math} in the electro-weak 
interaction sector and the quarks \begin{math} (u,d,c,s,t,b) \end{math}  
in the strong interaction sector. For the model to hold true and also 
give the observed {\em gravitational} as well as inertial masses,  
 i.e. \begin{math} \rho \approx \tilde{\rho }\end{math} without a factor of two, 
one must hold the Higgs field responsible for growing macroscopic 
gravitational mass as well as inertial mass on the elementary constituent 
particles. The gravitational implications of the Higgs nechanicsm of growing inertial 
and gravitational masses on elementary particles have yet to be fully explored.

\end{document}